# The Effect of Electrode Size on Memristor Properties: An Experimental and Theoretical Study

Ella Gale, Ben de Lacy Costello and Andrew Adamatzky
Unconventional Computing Group
University of the West of England
Bristol, UK

*Abstract- The width of the electrodes is not included in the current phenomenological models of memristance, but is included in the memory-conservation (mem-con) theory of memristance. An experimental study of the effect of changing the top electrode width was performed on titanium dioxide sol-gel memristors. It was demonstrated that both the on resistance, $R_{on}$, and the off resistance, $R_{off}$, decreased with increasing electrode size. The memory function part of the mem-con model could fit the relationship between $R_{on}$ and electrode size. Similarly, the conservation function fits the change in $R_{off}$. The experimentally measured hysteresis did not fit the phenomenological model's predictions. Instead the size of the hysteresis increased with increasing electrode size, and correlated well to decreasing $R_{on}$.*

## I. Introduction

In 1971, the number of fundamental circuit elements was increased to four with the prediction of the memristor, which relates charge, $q$, to magnetic flux, $\varphi$ [1]. The theory was first applied to a real world device in 2008, when Strukov et al reported the creation of a $TiO_2$ memristor [2]. This caused significant interest in the scientific community, as memristors are non-linear, possess a memory, require low operating power and are good candidates for neuromorphic computing.

One of the challenges in adapting these devices for use on an industrial scale is to understand how device properties relate to fabrication parameters. We focused on the solution processed $TiO_2$ sol-gel memristor [3], which has the advantages of ease of manufacturing, defect tolerance and compatibility with flexible substrates. The effect of choosing a different metal for the electrode on the device properties has already been investigated [4]. Here we investigate the effect of the electrode size.

Since the creation of the Strukov memristor, there have been three notable attempts to theoretically model real world memristor devices in order to better understand and control device parameters (not including the many simulation papers which offer incremental improvements on the base theory). The first was Strukov's phenomenological model, used to model the Strukov memristor [2] and since successfully applied to several other memristor systems, such as [5]. The second was Georgiou et al's rewrite of Strukov's model as a set of Bernoulli equations that could be analytically solved to estimate memristor hysteresis based on input waveform [6]. The third was Gale's mem-con memristor model, which is built on different principles to Strukov's in being derived from magnetostatic theory rather than experimental observation. Essentially, Strukov and Gale's models offer two differing theories on memristor operation. As a result of these models' different approaches, their predictions on electrode size effects differ and thus an experimental study can highlight which theoretical model is more useful.

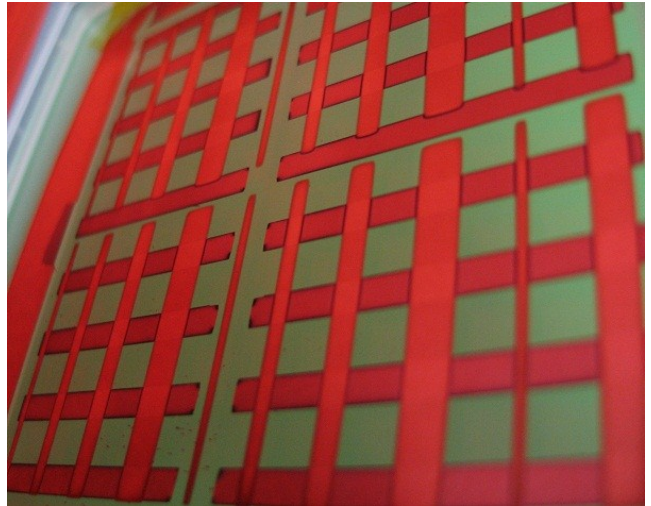

Fig. 1. Sheet of solution processed $TiO_2$ sol-gel memristors, with different sized electrodes. Note that the Aluminium electrodes are highly reflective and have been photographed under red light.

In this paper we report the experimental effects of changing the electrode size on the operation of the sol-gel memristor and the implications of this for memristor theory.

## II. $TiO_2$ Sol-Gel Memristor

The $TiO_2$ sol-gel memristor is a crossed-electrode device fabricated as described in Ref. [4] with the single difference that the mask used for the deposition of the top electrode has spacing of different widths. Fig. 1 shows the sheet of sputtered memristors with top electrodes of various sizes. The thickness of the $TiO_2$ sol-gel layer, $D$, is 40nm, the bottom electrode width, $E$, is 4mm and the top electrode width, $F$ was set to 1, 2, 3, 4 and 5mm widths.

The memristor works by the interconversion of $TiO_2$, the high resistance phase of resistance $R_{off}$, to the doped form $TiO_{(2-x)}$, the low resistance phase of resistance $R_{on}$. Although the picture may be more complex (see for example [7]) this model of the mechanism works well to describe the memristor's operation. As the oxygen vacancies move, the boundary, w, between the two types of material moves, changing the relative proportion of both and thus changing the memristor's resistance.

Reference [2] reported two different types of memristors from the same fabrication process, 'triangular' memristors which switched over a small voltage range and 'curved' memristors which switched between resistance states at a more slow and regular rate. It was noted in that paper that devices could be classified by doing a small-scale ±0.5V I-V curve. Devices that were ohmic over this range tended to be 'triangular' memristors over a larger range, those which possessed a distinctive open curve, similar to that seen before for $TiO_2$ [8], tended to be 'curved' types. In this work, I-V curves of ±0.5 V were run for devices of different electrode size.

III. THEORETICAL BACKGROUND

The definition of a memristor [1] relates the change in magnetic flux, $\varphi$, to the change in charge, $q$, within the device:

$$d\varphi = M(q)\, dq . \quad (1)$$

A. *Strukov's Phenomenological Model*
Starting from a description of the movement of the boundary between doped and undoped $TiO_2$, w, Strukov's model [2] gives the value of the Memristance, $M(q)$, as

$$M(q) = R_{off}\left(1 - \frac{\mu_v R_{on}\, q(t)}{D^2}\right), \quad (2)$$

where $\mu_v$ is the ion mobility of the oxygen vacancies, $q$ is the charge and $D$ is the thickness of the device. Note that the memristance and thus the I-V curves of the device depend only on these parameters, and therefore the electrode widths $E$ and $F$, have no effect on the memristance.

B. *Georgiou's Bernoulli equations*
Strukov's model was rewritten as Bernoulli equations primarily to present a method for predicting the size and shape of the memristor current response to different voltage waveforms [6]. All relevant physical dimensions of the device were combined into the 'dimensionless lumped parameter', $\beta$, and its rescaled version, $\tilde{\beta}$, as given by

$$\tilde{\beta} = 2\beta = \frac{2\,V_{max}}{\omega_0 R_0^2}\mu_v\left(\frac{R_{on}}{D}\right)^2\left(\frac{R_{off}}{R_{on}} - 1\right), \quad (3)$$

where the maximum voltage, $V_{max}$, the frequency the I-V curve is run at, $\omega_0$, and the starting resistance, $R_0$, have been included as physical parameters to be considered by the theory. This allowed the authors to give analytical expressions for the scaled hysteresis, $\bar{H}$, (scaled relative to $R_0$) for two waveforms, bipolar piecewise linear (BPWL) and triangular. These analytical solutions are above and below the numerically simulated value for a sinusoidal waveform (which is not analytically solvable) and as such give upper and lower bounds to the sinusoidal hysteresis.

C. *Mem-Con Theory*
Gale [9] starts from electrodynamics and derives the memristance by calculating the magnetic flux that arises from the vacancy movement. This gives a fundamentally different value for the memristance which is

$$M(q) = U\, X\, \mu_v\, P_k(q(t))\, , \quad (4)$$

where $U$ is the universal constants, $\mu_0/4\pi$, and $\mu_0$ is the permittivity of a vacuum, $X$ are the experimental constants given by the product of the area of one side of the device and the applied electric field. The only term that varies as the device charges is $P_k$ which arises from the magnetic field and is a function of $D$, $E$, $F$ and $w$. Only this theory of memristance requires knowledge of all three dimensions of the device and thus a comparison of the effect of electrode size on memristance will lead to a differentiation between the three theoretical models.

The mem-con theory requires that the memristance is fit experimentally to the memory function, $M_e$, given by

$$M_e = C_M * M + C_2, \quad (4)$$

where $C_M$ and $C_2$ are experimentally determined constants. The memristance relates the charge and flux associated with the oxygen vacancies. As the memory function has to be expressed in terms of the conducting electrons, the memristance must be fit using experimental data.

The conservation function, $R_{con}$, comes from the conservation of volume, i.e. that the volume of undoped $TiO_2$ shrinks as the volume of doped $TiO_{(2-x)}$ increases. This is given by

$$R_{con} = \frac{(D - w(t))\rho_{off}}{E\,F} \quad (5)$$

and this also includes all three spatial dimensions of the device. The total resistance is a sum of the memory and conservation functions, hence the name of this theory.

III. METHODOLOGY

A. *Experimental*
I-V curves were run with Keithley 2400 Sourcemeter, using sinusoidal voltage waveforms with an amplitude of 0.5V and a frequency of 0.68Hz.

B. *Theoretical*
We looked for an effect of the top electrode size on the maximum and minimum measured resistances. Gale's theory predicts an effect, Strukov's doesn't. To test Georgiou's theory, we calculated $\tilde{\beta}$ and $\bar{H}$ for the memristors. Georgious's theory predicts a relationship between the two variables and that the scaled hysteresis for the sinusoidal waveform should be between the predicted scaled hysteresis for the BPWL and triangular waveforms.

## IV. RESULTS AND DISCUSSION

### A. Effect of Electrode Size on the I-V Curves

59 memristors were run and classified based on the shape of their I-V curve. 33 exhibited ohmic behavior and 16 were curved devices. There were also 3 with circular open loops, 4 that exhibited triangular switching over this range, 3 that exhibited curved switching and 3 that were not connected.

For the triangular memristive switches there was no correlation between the electrode size and the resistance of the device. This suggests that these operate via a filamentary mechanism as the connected filament is a local effect and not a bulk effect (we would expect a bulk effect to be related to the bulk volume). There was also no noticeable correlation between the types of device and the electrode size, i.e. controlling this fabrication parameter does not provide a route to selecting device properties.

The open loop is shown in Fig. 2. There is an peculiarity of the negative current seen at positive voltage and vice versa. We suspect that this is related to the inertia of the moving oxygen ions. Similar effects have been seen in experimental flux-controlled memristor models [10]. Because [10] presented experimental results from a circuit made with typical electronic components (i.e. not ionic ones) selected to model a memristor according to Chua's definition, it suggests that this effect is part of the memristive action rather than a corollary to it.

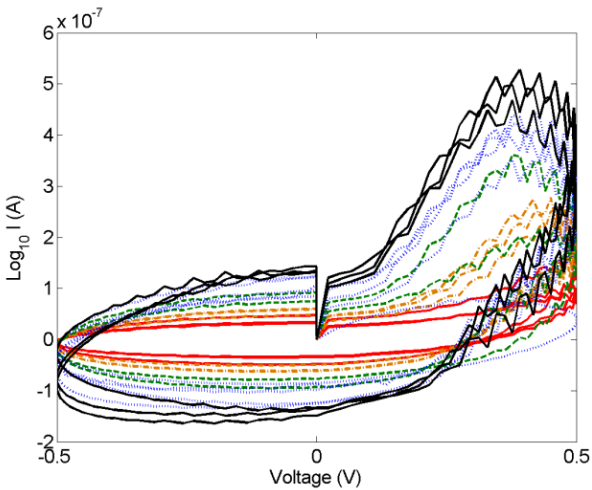

Fig. 2. I-V curves for the open-loops seen for curved memristors. Solid red: 1mm, Orange dot dashes: 2mm, Green dashes: 3 mm, Blue dots: 4mm and solid black 5mm.

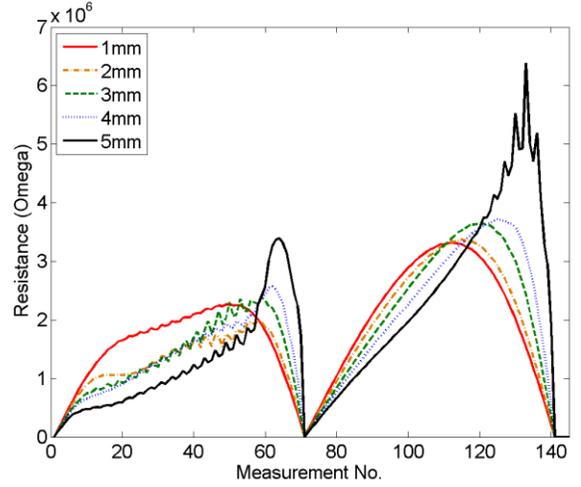

Fig. 3. The resistance profiles over the I-V loop. Both the size of the hysteresis and the lag increases with electrode size.

Of the 16 curved devices, there were 5 with $F$=4mm, 2 with $F$=3mm and 3 each of $F$=1mm, $F$=2mm, and $F$=5mm. The I-V curves are shown Fig. 2. There is a clear increase in hysteresis with electrode size. Note that there is one 4mm line which is much smaller than the rest, this outlier can be seen in all of the data. As this data point fits in well with the data for thinner electrodes, we suspect that this device had a cracked electrode.

Fig. 3 shows the averaged absolute value of resistance for these devices. The asymmetry is a measure of the lag due to the ion movements. The size of the peaks and the asymmetry are a measure of the hysteresis.

### B. Testing the Mem-Con Model of Memristance

As Strukov's theory, and as a result Georgiou's, is one-dimensional and only depends on $D$, there is no predicted change to the memristance as a result of different electrode sizes.

The mem-con theory predicts a difference in memristance as a function of electrode size. Both the memory and conservation functions decrease with electrode width. The experimental value for $R_{on}$ is the smallest resistance measured over the run. This is not the limit of $R_{on}$ because we do not know if we have fully discharged the device, however, we can use this minimum value as an approximation for it. The theoretical value of $R_{on}$ is calculated by finding the value for the memory function at the limit as $w \rightarrow D$, i.e. when the whole device is TiO$_{(2-x)}$.

We can use the memory function to fit the minimum resistance (as the memory function is the dominant term as $w \rightarrow D$ and $R_{Tot} \rightarrow R_{on}$). This gives the graph shown in Fig. 4 and it can be seen that the memory function fits the effect of electrode size over range of $R_{on}$ well. Note that this has been done with 2 experimentally determined parameters whose values are $C_M = 2.09 \times 10^{25}$ and $C_2 = 1.25 \times 10^{-20}$. As $C_2$ is so small compared to the range of $R_{on}$, we believe that one experimentally determined variable is enough. We expect $C_M$ to be related to the device physics and a theoretical basis for it is currently under investigation.

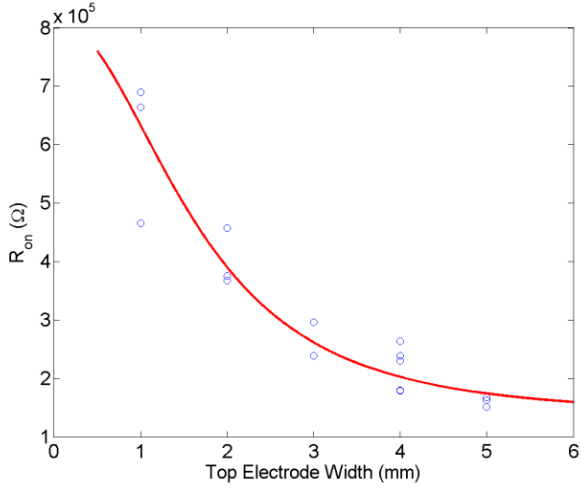

Fig. 4. Experimental values for $R_{on}$ fit by the Memory function from the mem-con model.

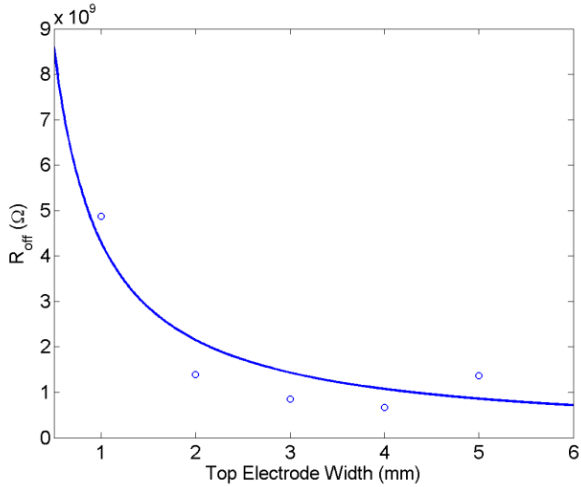

Fig. 5. Experimental values for $R_{off}$ fit by the Conservation function from the mem-con model.

For these results, the other two theoretical models predict no change in $R_{on}$. This validates the mem-con approach and highlights that the memristance is best thought of as a three-dimensional parameter. (This has been discussed for other memristors, where the perpendicular nature of the ion and electron current flows necessitates such a 3-D description, in [9].)

$R_{off}$ is not as highly correlated with electrode size as $R_{on}$, demonstrating that a 1-D model might work relatively well here. However, greater accuracy is gained by fitting the $R_{con}$ model to the experimental data (see Fig. 5). To obtain the experimental results, we measure the highest resistance state (this is visited three times after the start of the run and always has the same value for a particular device). As there was greater variance in the $R_{off}$ values, the averages for each electrode width are plotted. To get the theoretical numbers, we used the value of $R_{con}$ as $w \rightarrow 0$, i.e. when the whole device is TiO$_2$. The resistivity $\rho_{off}$ was used as a fitting parameter (where $\rho_{off} = 6.82 \times 10^{10}$), which is reasonable as we do not exactly know which phase the device is in. The conservation function fits the data well. Note that the other theoretical models described here predict a constant value for all these devices, which is apparently not correct.

### C. Testing Georgiou's quantitative measure of hysteresis

The hysteresis of the device, $H$, is calculated as in Ref. [6] and is a measure of the work taken to go round the loop, as calculated by the difference in work between the lower branch of the I-V curve and the upper one. The scaled hysteresis is this value divided by the work taken to drive a resistor of $R_0$ at the same voltage waveform and frequency.

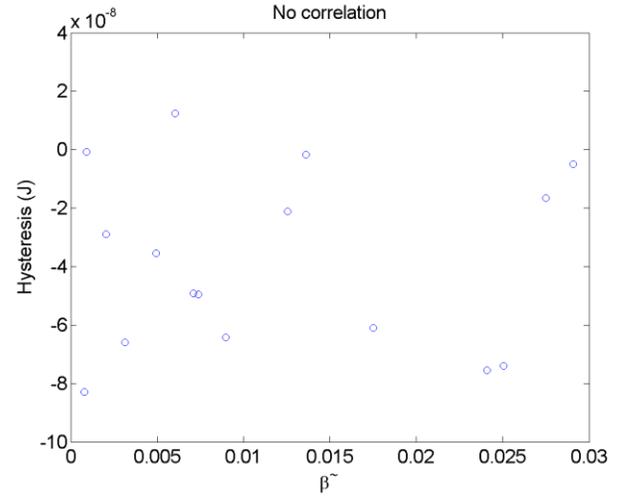

Fig. 6. The experimental results do not show any correlation between the hysteresis and the lumped dimensionless parameter $\tilde{\beta}$, contrary to the theoretical predictions in ref [6].

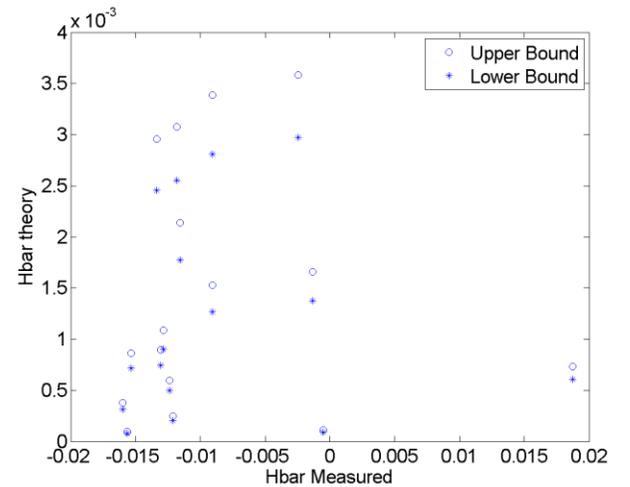

Fig. 7. There is no correlation between the theoretical upper and lower bounds for $\bar{H}$ (scaled hysteresis) and the experimentally measured $\bar{H}$, contradicting the theoretical predictions in [6].

Contrary to Georgiou's theoretical prediction we did not observe a correlation between $\tilde{\beta}$ and either the hysteresis, $H$, or the scaled hysteresis, $\bar{H}$ (see Fig. 6).

As $\bar{H}$ for bipolar piecewise linear waveforms (with $m=20$, see [6]) and triangular waveforms offer an upper and lower limit on the value of $\bar{H}$ for a sinusoidal waveform, we compared these theoretical limits with the experimentally calculated scaled hysteresis and determined that there was a difference of ~3 orders of magnitude between the theoretical and experimental values. As shown in Fig. 7, there was no correlation between Georgiou's theoretical predictions and the experimental values.

Our calculation of the hysteresis gives a negative number as the upper branch (the on state) has a larger hysteresis with respect to a resistor of resistance $R_0$ than the lower one does, except for the outlier.

From this data we are forced to conclude that the Bernoulli equations based on Strukov's model do not work for these memristors, although it is possible that this theory will be of use in predicting responses from different types of memristor. We cannot speculate from these data whether the Bernoulli equation approach would work if applied to the mem-con theory.

*C. Which Device Properties Cause the Change in Hysteresis Size?*

If the hysteresis does not depend on the lumped dimensionless parameter $\tilde{\beta}$, contrary to the statements in [6], then what does it depend on? From experimental data we have found two possibilities. As the top electrode size increases, the hysteresis shrinks, as shown in Fig. 8. This can be fit to a straight line with the equation

$$H = m_f F + c_f , \qquad (6)$$

where $m_f = -1.68 \times 10^{-8}$, $c_f = 1.39 \times 10^8$ and the norm of the residuals is $7.18 \times 10^{-8}$.

The hysteresis size is also negatively correlated with $R_{on}$ (there is no such correlation with $R_{off}$). This data is best fitted to a straight line if we use the logarithm of the hysteresis, as:

$$\log_{10}(H) = m_R R_{on} + c_R , \qquad (7)$$

where $m_R = -3.39 \times 10^{-6}$, $c_R = -6.55$ and the norm of the residuals is 0.845, see Fig. 9.

Whether the electrode size causes the change in hysteresis size directly or via changing the $R_{on}$ is not known, but there are a few facts that suggest the latter. $R_{on}$ is correlated with electrode size thus:

$$R_{on} = m_g F + c_g , \qquad (8)$$

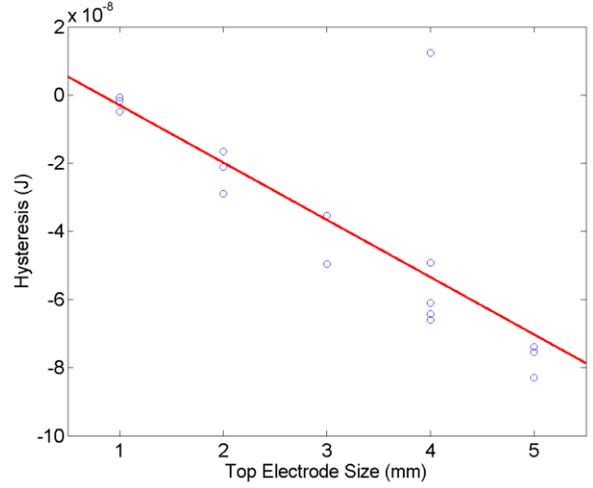

Fig. 8. The hysteresis increases with electrode size. The magnitude indicates the size of the hysteresis, the sign indicates that that the hysteresis is not equal but there is more in the top branch. Note that the outlier visible in Fig. 1 is the only device with a positive hysteresis.

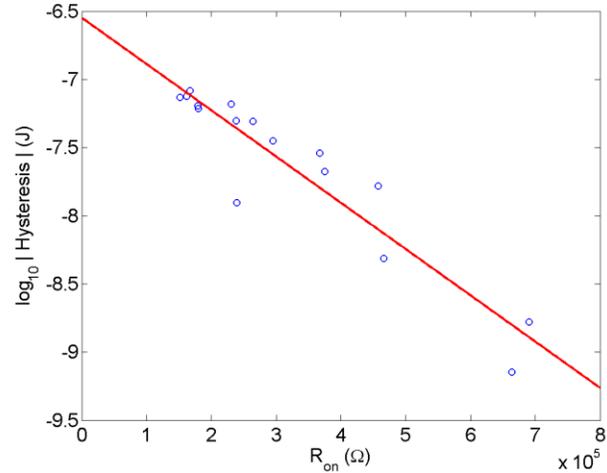

Fig. 9. The hysteresis is related to the measured $R_{on}$.

where $m_g = -1.08 \times 10^5$, $c_g = 6.58 \times 10^5$ and the norm of the residuals is $2.68 \times 10^5$. The hysteresis is a measure of the interaction of two sets of parameters. The upper and lower limits of the current, and thus the limits of the loop are prescribed by $R_{off}$ and $R_{on}$. These maximum and minimum resistances for the fully switched device are $R_{off}|_{w \to 0}$ and $R_{on}|_{w \to D}$, respectively. The interaction between $\omega_0$ and $\mu_v$ affect the amount $w$ moves, and thus the value of $R_{off}$ and $R_{on}$ compared to these limits. Therefore, a fabrication parameter that changes the value of $R_{on}$ would be expected to change the value of the hysteresis. Note that there is no correlation between the ratio $R_{on}$ to $R_{off}$ and $H$, neither is there a correlation between $R_0$ and this ratio.

Fig. 9 shows an increase in the size of the hysteresis with increasing Ron, and the outlier point is not an outlier here. This suggests that $R_{on}$ is the better fabrication parameter to use for predicting the hysteresis and also that the outlier device had a cracked electrode so that the effective top electrode width was less than the 4mm it should have been.

V. CONCLUSIONS

Three theories of memristance have been compared. The mem-con theory correctly predicts that $R_{off}$ and $R_{on}$ will both decrease with increasing electrode size. When $w \to 0$ the memristance is entirely described by the memory function and this function fit the relation between the measured $R_{on}$ and $F$ very well. Similarly, as $w \to D$ the memristance is entirely described by the conservation function and this function was found to fit the relation between $R_{off}$ and $F$ with only one fitting parameter.

Georgiou's lumped parameter did not accurately predict the size of the hysteresis. Strukov's one-dimensional model did not predict any effect of changing electrode width, as this factor was not included within the model.

We have demonstrated that changing the size of an electrode affects the behavior of curved type memristors and has no effect on triangular switching ones. This suggests that the two types operate via different mechanisms. The size of the hysteresis increases with increasing electrode size, as a result of the decrease in the value of $R_{on}$ with increasing electrode size.

The experimental results presented in this paper suggest that that a three-dimensional model of memristance is needed and that the Mem-Con model [9] gives a good fit to the experimental data.

ACKNOWLEDGMENT

There will be something here.